\documentstyle[twoside,fleqn,epsfig,amssymb,amsmath,espcrc2,graphicx]{article}

\newcommand{\msbar}{{\overline{\rm MS}}}
\newcommand{\bea}{\begin{eqnarray}}
\newcommand{\eea}{\end{eqnarray}}
\newcommand{\beq}{\begin{equation}}
\newcommand{\eeq}{\end{equation}}
\newcommand{\gev}{{\rm GeV}}

\newcommand{\pdir}{p\kern -5.2pt\raise 0.2ex\hbox {/}}
\newcommand{\vdir}{v\kern -5.75pt\raise 0.15ex\hbox {/}}
\newcommand{\kdir}{k\kern -5.75pt\raise 0.15ex\hbox {/}}
\newcommand{\epsdir}{\epsilon\kern -5.0pt\raise 0.15ex\hbox {/}}
\newcommand{\bvdir}{\bar{v}\kern -5.75pt\raise 0.15ex\hbox {/}}
\newcommand{\Ddir}{D\kern -7.75pt\raise 0.20ex\hbox {/}}
\newcommand{\ldir}{l\kern -5.0pt\raise 0.2ex\hbox{/}}
\newcommand{\varepsdir}{\varepsilon\kern -5.5pt\raise 0.15ex\hbox{/}}

\newcommand{\kkbar}{K^0-\bar K^0}

\title{Matrix elements of $\Delta S=2$ operators with Wilson fermions}

\author{D.~Becirevic\address{Dip. di Fisica, 
Universit\`a ``La Sapienza" and INFN-Roma, P.le A. Moro 2, I-00185 Rome,
Italy}, Ph.~Boucaud\address{Laboratoire de Physique Th\'eorique (b\^at.210),
Universit\'e de Paris XI, 91405 Orsay Cedex, France}, 
V.~Gim\'enez\address{Dep.~de F\'{\i}s.Te\`orica and IFIC, Univ.~de Val\`encia, Dr.~Moliner 50, E-46100, Burjassot, Val\`encia,
Spain}, 
V.~Lubicz\address{Dip. di Fisica, Univ. di Roma Tre and INFN - Roma Tre, 
Via della V. Navale 84, I-00146 Rome, Italy},   
G.~Martinelli$^{\rm a}$, M.~Papinutto\address{Dip. di Fisica, Univ. di Pisa 
and INFN - Pisa, 
Via Buonarroti 2, I-56100 Pisa, Italy}\thanks{
Talk given by Damir Becirevic.}}
\begin{document}
\newcommand{\sze}{\small}

\begin{abstract}
We test the recent proposal of using the Ward identities 
to compute the $\kkbar$ mixing amplitude  
with Wilson fermions, without the problem of
spurious lattice subtractions. From our simulations, we observe no  
difference between results obtained with and without subtractions. 
From the standard study of the complete set 
of $\Delta S=2$ operators, we quote the following (preliminary) 
results: 
$B_K^\msbar (2 \ GeV)=0.70(10)$, 
$\langle O_{7}\rangle_{K\to \pi \pi}^{(I=2)} = 0.10(2)(1) \gev^3$, 
$\langle O_{8}\rangle_{K\to \pi \pi}^{(I=2)} = 0.49(6)(0) \gev^3$.
\end{abstract}
 
\maketitle
The main problem in lattice computations of the 
$4$-fermion $\Delta F=2$ operators with Wilson fermions 
is related to spurious mixing among dimension-six operators. 
For example, chiral symmetry ensures that 
the operator responsible for the indirect $CP$-violation in the 
$\kkbar$ system, $Q_{1} = \bar s^a \gamma_\mu (1 - \gamma_5)d^a  
\ \bar s^b \gamma_\mu (1 - \gamma_5)d^b$, renormalizes multiplicatively. 
Since the Wilson term explicitly breaks chiral symmetry, 
$Q_1$ mixes instead with all the other $\Delta S =2$ operators including
\bea \label{SUSY}
&& Q_{_2} = \bar s^a  (1 - \gamma_5)d^a  \ \bar s^b  (1 - \gamma_5)d^b\;\cr
{\phantom{\Huge{l}}}\raisebox{-.2cm}{\phantom{\Huge{j}}}
&& Q_{_3} = \bar s^a  (1 - \gamma_5)d^b  \ \bar s^b (1 - \gamma_5)d^a\;\cr
{\phantom{\Huge{l}}}\raisebox{-.15cm}{\phantom{\Huge{j}}}
&& Q_{_4} = \bar s^a  (1 - \gamma_5)d^a  \ \bar s^b (1 + \gamma_5)d^b\;\cr
{\phantom{\Huge{l}}}\raisebox{-.15cm}{\phantom{\Huge{j}}}
&& Q_{_5} = \bar s^a  (1 - \gamma_5)d^b  \ \bar s^b (1 + \gamma_5)d^a\;
\eea 
where the superscripts denote color indices. 
The spurious mixing may seriously modify the chiral 
behavior of the operator $Q_1$ and hence need to be subtracted away,  
$Q_1^{(sub.)}(a) = Q_1(a) + \Delta_i(a) Q_i(a)$ $(i=2,\dots,5)$.
After completing the subtraction procedure of the 
``wrong chirality" operators, the bare lattice regularized operator $Q_1^{(sub.)}(a)$ must 
be multiplicatively renormalized (like in the continuum): 
$\hat Q_1(\mu) = Z_{11}(\mu a) Q_1^{(sub.)}(a)$. 
A complete programme of renormalization of the operators $Q_{1-5}$ 
requires the computation of 9 renormalization and 16 subtraction constants.
This can be done perturbatively (see ref.~\cite{lanl}), but since the 
${\cal O}(\alpha_s)$ terms are uncomfortably large, a non-perturbative
determination of these 25 constants is mandatory. A theoretically simple
method to implement the renormalization in the so-called (Landau)RI/MOM scheme has been 
summarized in ref.~\cite{bibbia}. In practice this programme is, 
however, quite complicated. Even more so since the procedure 
included also the necessity to subtract the Goldstone boson 
(single and double pole) contributions~\cite{alain,dawson,mauro}. 
That may cast doubts on the
reliability of the method, {\it i.e.} on the level of control of the 
systematic errors.
To address that issue, one needs a method 
allowing a computation without the necessity to subtract 
mixing with operators of the wrong chirality,  
and compare the results to the ones obtained by using 
the standard method (with subtractions). 
Recently, two such proposals appeared: 
twisted mass QCD~\cite{sint} (see \cite{pena} for the first numerical results), 
and method of the Ward identities~\cite{mi} which 
we use is what follows.
\section{Ward Identity Method~\cite{mi}}
The method is extremely simple and to summarize it in a few lines 
we write the parity even (PE) operator $Q_1$ as $Q_1 = VV + AA$, in an 
obvious notation. 
For symmetry reasons, unlike the PE, the 
parity odd (PO) operators do not suffer from spurious mixings 
({\it i.e.}  $\Delta_{ij}^{PO}(a)=0$). It is then possible to apply 
the Ward identity on the matrix element of the PO operator, 
${\cal Q}_1 = AV + VA$, to get the matrix element of the PE one, $Q_1$. 
By applying the chiral rotation around the 
third axis in isospace, $\delta u  = \gamma_5  u$, $\delta d  =  - \gamma_5  d$ ($m_u = m_d \equiv m$),  
on  $\langle  P(\vec x,t_x)  {\cal Q}_1 (0)  P(\vec y,t_y) \rangle$, one arrives at
\bea
&&\hspace*{-7mm} m \langle \displaystyle{\sum_{\vec x,\vec y,\vec z,t_z}} \Pi (\vec z,t_z)\
 P(\vec x,t_x)\  {\cal Q}_1 (0) \
 P(\vec y,t_y) \rangle =\cr
&&\hspace*{-7mm}   \langle \displaystyle{\sum_{\vec x,\vec y}}
P(\vec x,t_x)\  Q_1 (0) \ P(\vec y,t_y) \rangle
+ {\rm (rot.\ sources)}
\eea 
where the terms arising from the rotation of the source operators 
$P = \bar d \gamma_5 s$, vanish due to the symmetry under the charge 
conjugation. In the above identity, $\Pi = \bar d \gamma_5 d  - \bar u \gamma_5 u$. 
Thus, to get an informations on the r.h.s., we compute the l.h.s. where the operator 
${\cal Q}_1$ renormalizes only multiplicatively. Moreover, $Z_P(\mu)$, which renormalizes the
density $\Pi$, cancels against the one appearing in the mass renormalization constant, $Z_m(\mu) =
Z_A/Z_P(\mu)$, so that only $Z_A$ is required. 
We computed the l.h.s. of the above identity on the lattice at 
$\beta =6.0$ ($16^3\times 52$) and at $\beta =6.2$ ($24^3\times 64$). 
Results of both simulations were presented at the conference. Due to the lack of space, here we 
present only the results obtained on the finer lattice ($\beta =6.2$, 200 configurations), where 
we work with $\kappa_q\in \{ 0.1339, 0.1344, 0.1349, 0.1352\}$. Complete results with 
accompanying details will be presented in ref.~\cite{damir}.

To confront the two methods, we studied  the following ratios:
\bea
&&\hspace*{-7mm}{\cal R}_1(t_x) = { m    \ \langle
\displaystyle{\sum_{\vec x,\vec y,\vec z,t_z}} \Pi (\vec z,t_z)\
 P(\vec x,t_x)\ \hat {\cal Q}_1 (0) \
 P(\vec y,t_y) \rangle \over Z_A\ \langle  \displaystyle{\sum_{\vec x} }
P (\vec x,t_x) P (0) \rangle
\langle  \displaystyle{\sum_{\vec y}
P (\vec y,t_y) P (0)\rangle }  }\nonumber ,\eea \bea
&&\hspace*{-7mm}{\cal R}_2(t_x) ={\langle \displaystyle{\sum_{\vec x,\vec y} }
 P(\vec x,t_x)\ \hat Q_1 (0) \
 P(\vec y,t_y) \rangle \over Z_A^2\ \langle  \displaystyle{\sum_{\vec x} }
P (\vec x,t_x) P (0) \rangle
\langle  \displaystyle{\sum_{\vec y}
P (\vec y,t_y) P (0)\rangle }  } ,\nonumber 
\eea                
where, when needed, we used the non-perturbatively determined 
renormalization and subtraction constants. In the above formulae the time $t_y$ has been 
fixed, $t_x$ is left free and $t_z$ has been  summed over all lattice. The role of denominators is to
eliminate the usual exponential terms in the numerator.

\begin{figure}[htb] \vspace*{-7mm}
\hspace*{-4mm}\includegraphics[width=18pc]{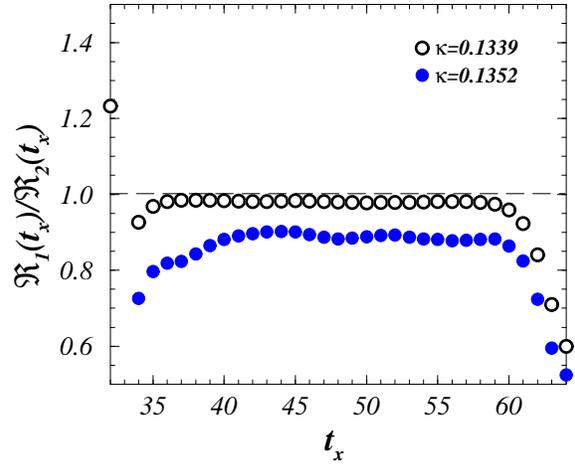}\vspace*{-8.5mm}
\caption{\sl \small Method without subtractions (${\cal R}_1(t_x)$) is confronted to the standard one -- 
with subtractions (${\cal R}_2(t_x)$). Subtraction constants are computed non-perturbatively.}
\label{fig1}\vspace*{-7mm}
\end{figure}
\begin{figure*}[t!] 
\vspace*{-7cm}
\includegraphics[width=18pc]{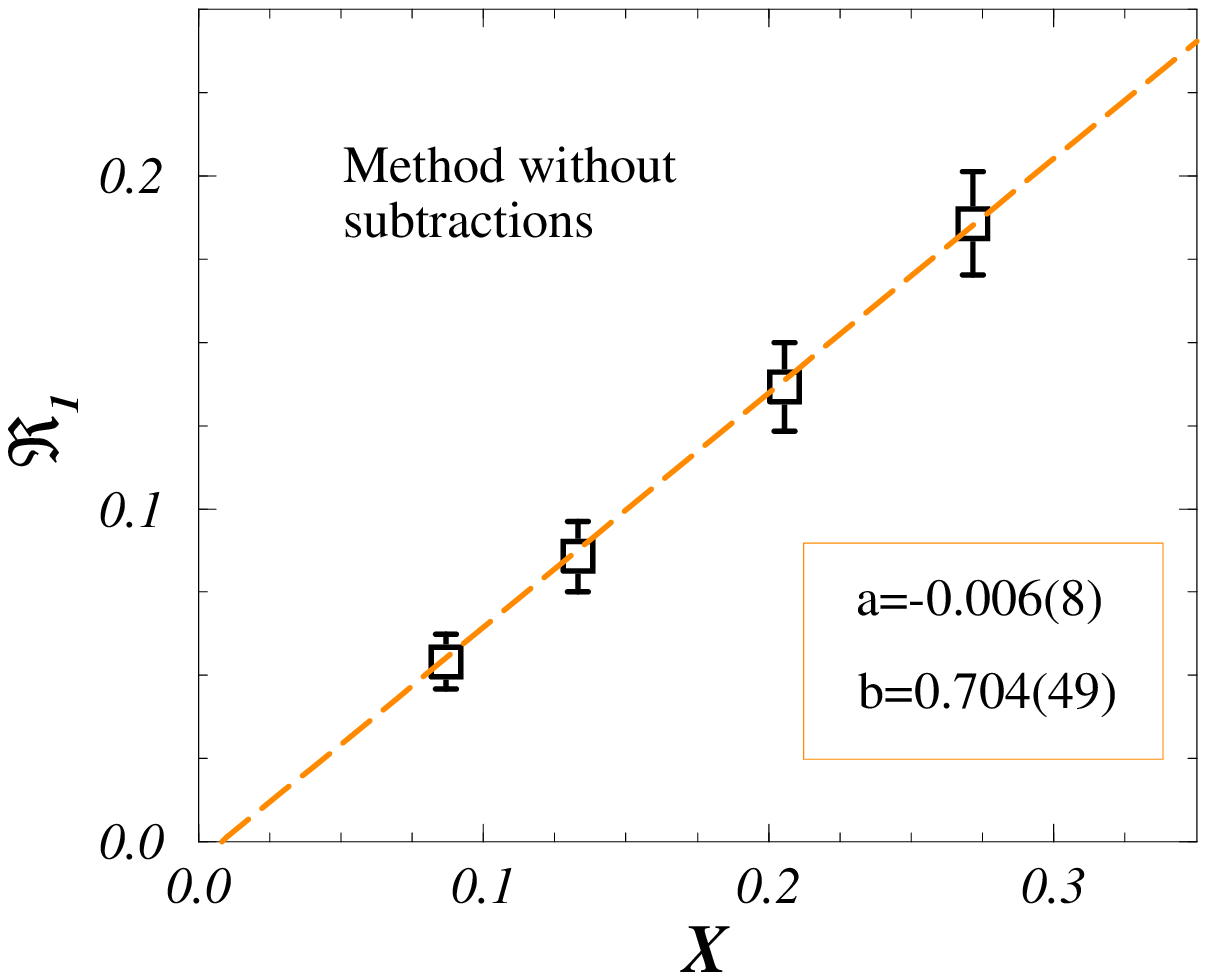}
\includegraphics[width=18pc]{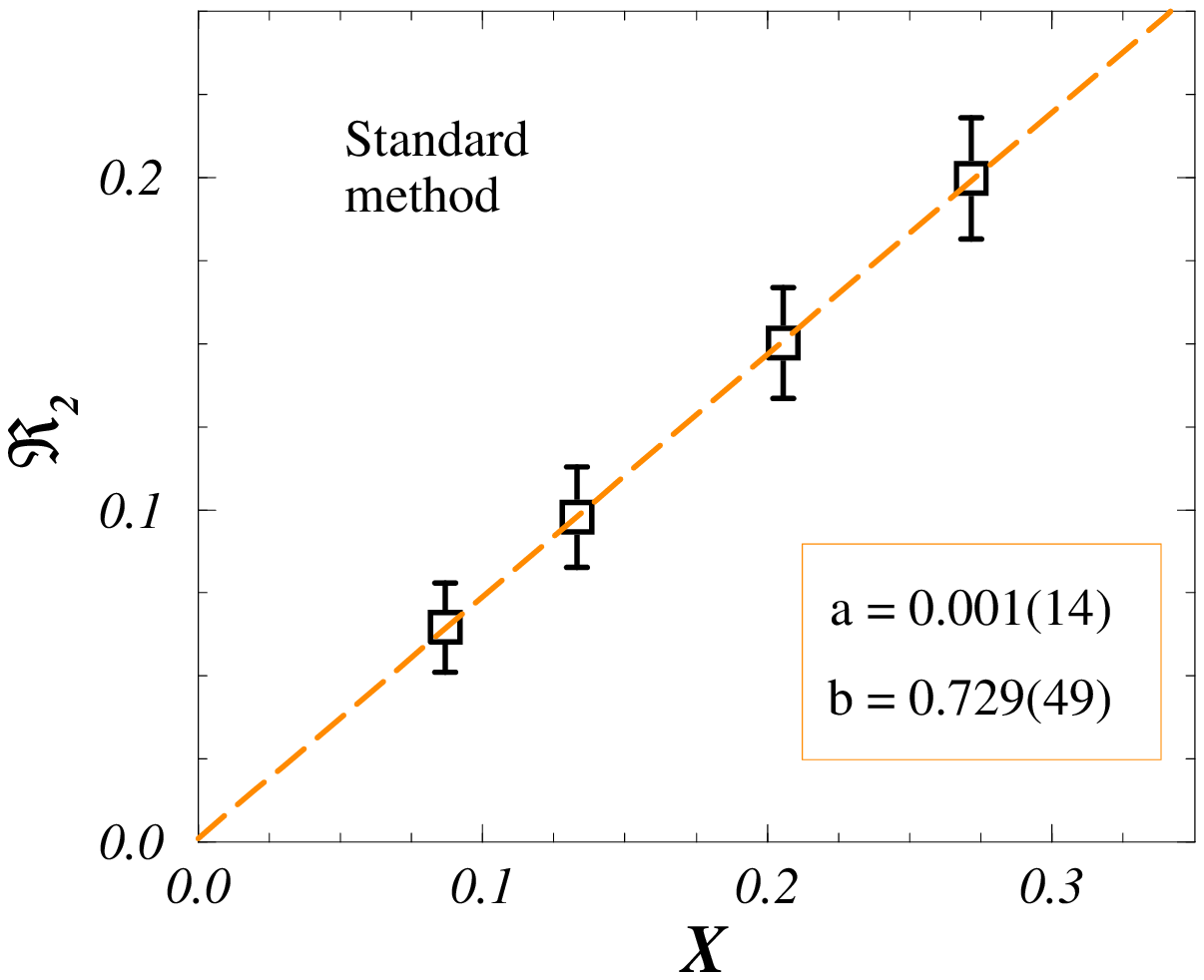} \\
\caption{\sl \small Chiral behavior of the matrix element of the operator 
${\hat Q}_1(\mu\approx 1/a)$ as obtained by using the
 methods without (left figure) and with subtractions (right figure).
The values of the fit parameters ${\cal R} = a+ b X$ are also given ($X\propto 
 m_q$, see ref.~\cite{add} for the precise definition).}
\label{fig2}\vspace*{-7mm}
\end{figure*}

                                                                    
First, we observe that the 
plateaus for the ratio ${\cal R}_1(t_x)$ exist and they are good for all the quark masses 
used in our study (see~\cite{damir}). Second, by using the method without 
(${\cal R}_1(t_x)$) and with (${\cal R}_2(t_x)$) subtractions, we get the results very 
consistent with each other. 
In fig.~\ref{fig1} we show the ratio ${\cal R}_1/{\cal R}_2$ for the smallest and the largest 
of our quark masses. From the observation that (on the plateau) for all our masses we get 
$0.89 \leq {\cal R}_1/{\cal R}_2 \leq 0.98$, we
conclude that {\bf the agreement of the two methods is satisfactory}. Note that results of
two methods have different $O(a)$ effects. 
We checked that the chiral behavior for the matrix element $\langle Q_1(\mu)\rangle$, as extracted by
using either of the two methods, is good, {\it i.e.} that ${\cal R}_2(t_x) \to 0$ when $m_q \to 0$ (see
fig.~\ref{fig2}).
In addition, we performed the analysis of the data (w/o subtractions) when a small momentum 
is given to the external sources (``kaons"). After following the usual extrapolation procedures~\cite{add},
we get\vspace*{-2mm}
\bea \label{bk}
&&\hspace*{12mm}B_K^{\msbar} = 0.70 (10)\,.  \eea 
This error can be substantially reduced if the computation is made at several lattice spacing 
so that the momentum injections to the external kaons (for which the signals are noisier) 
are not needed~\cite{damir}.

\section{$B$-parameters of the SUSY operators}

As seen in the previous section, the two methods give very consistent results 
which makes us more confident that systematic errors introduced by using 
the standard method (with subtraction and renormalization constants 
computed non-perturbatively), are indeed under control. We used the standard method to compute 
the matrix elements for the SUSY operators listed in eq.~(\ref{SUSY}). They are parameterized as
\bea
&&\hspace*{-7mm}\langle \bar K^0 \vert \hat Q_{2-5}(\mu) \vert K^0 \rangle =
B_{2-5}(\mu) \  c_{2-5} \times \cr
&&\hspace*{+7mm} \times \langle \bar K^0 \vert  \bar s \gamma_5d  (\mu) 0 \rangle
\langle 0 \vert  \bar s \gamma_5 d (\mu) \vert K^0 \rangle \ , \eea
where $c_2=-5/3$, $c_3=1/3$, $c_4=2$, $c_5=2/3$, so that the corresponding $B$-parameters are unity in 
the vacuum saturation approximation. Each $B_i$-parameter is computed by replacing $Q_1\to Q_i$ in 
${\cal R}_2(t_x)$ and dividing by $Z_P^2(\mu)$ instead of $Z_A^2$. 
Conversion from the RI/MOM scheme to the $\msbar$(NDR) scheme of
ref.~\cite{buras} is made in perturbation theory with NLO accuracy. 
By linearly interpolating to the physical kaon mass, we obtain the following values:
\bea
&&B_2^\msbar (2\ \gev) = 0.64(5)(2)\,,\cr
&&B_3^\msbar (2\ \gev) = 0.97(8)(12)\,,\cr
&&B_4^\msbar (2\ \gev) = 0.87(6)(3)\,,\cr
&&B_5^\msbar (2\ \gev) = 0.58(5)(1)\,,
\eea
which, together with $B_1$ given in eq.~(\ref{bk}), gives a complete 
set of $B$-parameters of $\Delta S=2$ operators needed for the analysis 
of the SUSY effects in the $\kkbar$ mixing amplitude~\cite{luca}.

\section{Very briefly on $\langle (\pi\pi)_{I=2}\vert Q_{7,8}(\mu)\vert K^0\rangle$}

From the results of the previous section, one may get the useful information on 
the $\Delta I=3/2$ amplitude of the $K\to \pi\pi$ decay. After extrapolating to the chiral limit, 
the relevant bag parameters are: $B_5^\chi \equiv B_7^\msbar (2 \gev) = 0.46(6)(2)$, 
and $B_4^\chi \equiv B_8^\msbar (2 \gev) = 0.82(7)(3)$. 
To predict the matrix elements in physical units, we will use the 
recipe of ref.~\cite{leo}, which avoids the multiplication by the 
quark condensate (squared) and replaces it by the multiplication by $m_\rho^2 f_\pi^2$. 
Using that strategy, in the $\msbar$(NDR) scheme of
ref.~\cite{buras}, we obtain:
\bea
&&\hspace*{-5mm}\langle \pi^+ \vert O_7^{3/2}(2\ \gev)\vert K^+\rangle = - 0.0193(33)(13) \ \gev^4 , \cr
&&\hspace*{-5mm}\langle \pi^+ \vert O_8^{3/2}(2\ \gev)\vert K^+\rangle = - 0.092(10)(0) \ \gev^4 .\nonumber
\eea
These results, after using the soft pion theorem $\langle (\pi\pi)_{I=2}
\vert O_{7,8}\vert K^0\rangle =
- \langle \pi^+ 
\vert O^{3/2}_{7,8}\vert K^+\rangle/f_\pi\sqrt{2}$~\cite{claude}, 
lead to\vspace*{-.7mm}
\bea
&&\langle  O_7(2\ \gev)\rangle_{K\to \pi\pi}^{(I=2)} = 0.104(18)(7) \ \gev^3 \cr
&&\langle  O_8(2\ \gev) \rangle _{K\to \pi\pi}^{(I=2)}= 0.49(6)(0) \ \gev^3 ,\nonumber 
\eea
which can be compared to the preliminary estimates of the SPQcdR collaboration, 
$\langle  O_7(2\ \gev)\rangle_{K\to \pi\pi}^{(I=2)} = 0.021(11)~\gev^3$ and 
$\langle  O_8(2\ \gev)\rangle_{K\to \pi\pi}^{(I=2)} = 0.53(6)~\gev^3$, as obtained 
by directly computing $K\to \pi\pi$ matrix elements on the lattice. For comparison with 
other groups and other approaches, see ref.~\cite{guido}.
\noindent

\newpage
\end{document}